\documentclass[aip,jcp,preprint,endfloats]{revtex4-1}
\usepackage{graphicx}
\usepackage{color}
\usepackage{epsfig}
\usepackage{amsmath}

\begin{document}

\title{Driving Force of Water Entry in Hydrophobic Channels of Carbon Nanotubes: Entropy or Energy?}
\author{Hemant Kumar}
\email{hemant@physics.iisc.ernet.in}
\author{Chandan Dasgupta}
\email{cdgupta@physics.iisc.ernet.in}
\author{Prabal K. Maiti}
\email{maiti@physics.iisc.ernet.in}
\affiliation{Centre for Condensed Matter Theory, Indian Institute of Science, Bangalore, India-560012}
\date{\today}

\begin{abstract}
Spontaneous entry of water molecules inside single-wall carbon nanotubes (SWCNT) has been confirmed by both simulations and experiments. Using molecular dynamics simulations, we have studied the thermodynamics of filling of a (6,6) carbon nanotube in a temperature range from 273 K to 353 K and with different strengths of the nanotube-water interaction. From explicit energy and entropy calculations using the two-phase thermodynamics (2PT) 
method, we have presented a thermodynamic understanding of the filling behavior of a nanotube. We show that both the energy and the entropy of transfer decrease with increasing temperature. On the other hand, scaling 
down the attractive part of the carbon-oxygen interaction results in increased energy  of transfer while the entropy of transfer increases slowly with decreasing the interaction strength. Our results indicate that both energy and entropy favor water entry in (6,6) SWCNTs. Our results are compared with those of several recent
studies of water entry in carbon nanotubes.

\end{abstract}

\maketitle
\section{Introduction}
Water has been a subject of interest for researchers for many decades. The ubiquitous presence of water makes it an interesting molecule to study in various physical conditions. Understanding the characteristics of water in confined environment is important for designing various nanodevices~\cite{doi:10.1021/cr078140f,Holt19052006} as well as for understanding various biological activities associated with different protein channels~\cite{doi:10.1021/cr068037a,doi:10.1021/cr020661}. Both simulations~\cite{hummer_nature,koga_nature,biswa_acs} and experiments~\cite{koles_PRL,majumder_nature,PhysRevLett.110.157402,doi:10.1021/nn901554h,PhysRevE.85.031501} have confirmed spontaneous entry of water inside single-wall carbon nanotubes (SWCNTs). Inside the hydrophobic cavity of SWCNTs, water molecules exhibits many remarkable properties including  highly ordered structures~\cite{mann_prl,jcp2007}, anisotropic rotational dynamics~\cite{biswa_acs,jpcb2009} and fast flow rates~\cite{Holt19052006}. Recent studies have suggested potential applications of nanodevices based on carbon nanotubes for energy storage, desalination, voltage generation, flow sensing, fast flow devices etc~\cite{Ghosh14022003,doi:10.1021/jp405036c,book1}. Inside narrow SWCNTs, water molecules are arranged in a solid-like structure to form  single-file chains of molecules  with all dipoles pointing in the same direction along the nanotube axis~\cite{mann_prl,jcp2007}. This highly ordered one-dimensional structure of water molecules has been predicated to be thermodynamically stable~\cite{Kofinger09092008} for lengths up to millimeters at room temperature. Understanding the thermodynamics of water inside the hydrophobic cavity of a nanotube is interesting not only from the fundamental point of view but also to understand various biological activities where proteins undergo structural changes to control the presence of water inside various channels to perform specific activities~\cite{ISI:000238326400017,ISI:000255723500028,doi:10.1021/cr068037a}.\\

In spite of many extensive studies~\cite{hummer_nature,chandler_jpcb2003,zhou_jcp2004,hummer_jcp2004,hemant_jcp, pascal_pnas, waghe2012,garate}, the driving force of water entry into carbon nanotubes is not clear, mainly due to contradictory findings from different studies. Using molecular dynamics (MD) simulations, Hummer {\it et al.}~\cite{hummer_nature} first reported spontaneous entry of water molecules inside narrow SWCNTs and based on the energy distribution, proposed the driving force to be the gain in rotational entropy. Later Chandler~{\it et al.}~\cite{chandler_jpcb2003} used a one-dimensional lattice-gas model to explore the thermodynamics of the filling and emptying transitions of a hydrophobic tube. They explained the bimodal nature (empty and filled) of the state of a nanotube immersed in water from an analogy with the liquid-vapor transition where it is energetically unfavorable to create a liquid-vapor interface but the entropy gain compensates for the energy increase. This simple model explains various observations from detailed atomistic MD simulations. However, due to its approximate nature, this model has limited applicability. Later Vaitheeswaran~{\it et al.}~\cite{hummer_jcp2004} computed energy and entropy of transfer using the temperature dependence of the occupancy probability for periodically replicated short segments of (6,6) SWCNT and concluded favorable entropy of transfer and unfavorable energy of transfer for partial occupancy and the opposite behavior for fully occupied nanotubes. Recently, from energy and entropy calculations from atomistic MD simulation trajectories using the two-phase thermodynamics (2PT) method, we showed that water molecules inside a narrow SWCNT have higher rotational entropy as compared to the bulk and the gain in entropy is sufficient to compensate the increase of energy due to the loss of hydrogen bonding inside the cavity of the nanotube~\cite{hemant_jcp}. Pascal {\it et al.}~\cite{pascal_pnas} carried out a similar study for various nanotube diameters and found that the confined water molecules have higher entropy up to a critical diameter of $\sim 11$~\AA~ and beyond this critical diameter, the energy of the confined water molecules becomes favorable while the entropy decreases upon entry. Waghe~{\it et al.}~\cite{waghe2012} computed the free energy of transfer from the occupancy probability for reduced carbon-oxygen interaction and extrapolated it to the actual carbon-oxygen interaction to argue that the entry of water molecules is driven by a favorable energy of transfer. Recently Garate {\it et al.}~\cite{garate} studied the free energy 
of transfer for various occupancies of SWCNT using thermodynamic integration and showed that for partial occupancy, the entry is driven by entropy, whereas the entry is energy driven for higher occupancies.\\

It is important to understand the origin of such conflicting findings and to develop a clear understanding of the thermodynamics of water entry inside the cavity of carbon nanotubes. In this report, we study the thermodynamics of single-file water molecules confined in a SWCNT at various temperatures and  for different carbon-oxygen interaction parameters to develop a better understanding of the behavior. Based on these findings, we try to understand the origin of the discrepancies among the results of earlier studies.\\

The organization of the paper is as follows. In section~\ref{method}, we give the simulation details and then present the results for the free-energy of transfer at different temperatures in section~\ref{temp}. In section~\ref{vdw}, we study the thermodynamics of confined water for different carbon-oxygen interaction strengths.  In section~\ref{discuss}, we compare our result with those of the other studies and explain the origin of the discrepancies among them. 

\section{Method of Simulation}
\label{method}
To compute the free energy of filling, we have calculated the energy and entropy of water molecules in bulk and in confinement inside a SWCNT. The differences between the energy and entropy of bulk and confined water molecules give an estimate of the free energy of transfer. The energy of each water molecule can be directly computed from the simulation trajectories. To compute the entropy of water molecules, we have used the two-phase thermodynamics (2PT) method~\cite{Lin_JPC,Lin_JCP} based on the density of states which has been shown to give bulk water entropy values in excellent agreement with experiments~\cite{Lin_JPC}. This method also has been applied to study the entropy of water and several organic liquids under various conditions~\cite{jana_JPCB,ananya_jcp, ananya_prl,Spanu26042011,C0CP01549K}. Details of the method to compute the entropy using this scheme have been presented in our previous 
study~\cite{hemant_jcp}.\\

To study the thermodynamics of filling at various temperatures, we have performed a series of MD simulations for a 54~\AA~ long (6,6) open-ended armchair SWCNT immersed in a bath of bulk water ($\sim$ 5000 water molecules) at temperatures ranging from 273 K to 353 K. Interactions between various atom types were modeled using AMBER ff10 force field ~\cite{amber_ff}. Carbon atoms in the nanotube were modeled as  Lennard-Jones (LJ) particles without any charge (ff03 atom type ``CA''). We have used the TIP3P model ~\cite{tip3p} for water molecules and all OH bonds were constrained using the SHAKE algorithm. \\ 

To systematically explore the effect of the carbon-water interaction on the thermodynamics of filling, we performed the simulation with different carbon-oxygen interaction strengths. Following the procedure used by Waghe {\it et al.}~\cite{waghe2012}, we scaled the attractive part of the LJ interaction by a factor $\lambda$ such that:
\begin{equation}
U(r,\lambda)=4\epsilon\left[\left(\frac{\sigma}{r}\right)^{12}-\lambda\left(\frac{\sigma}{r}\right)^6\right]=4\epsilon^{'}\left[\left(\frac{\sigma^{'}}{r}\right)^{12}-\left(\frac{\sigma^{'}}{r}\right)^6\right]
\label{ljmod}
\end{equation}
where $\epsilon^{'}=\epsilon\lambda^2$ and $\sigma^{'}=\sigma/\lambda^{1/6}$ are the modified LJ parameters. We have studied the thermodynamics of filling for four different values of $\lambda=0.752,0.785,0.90,1.00$. %
\begin{figure}
\centering
\includegraphics[scale=0.8]{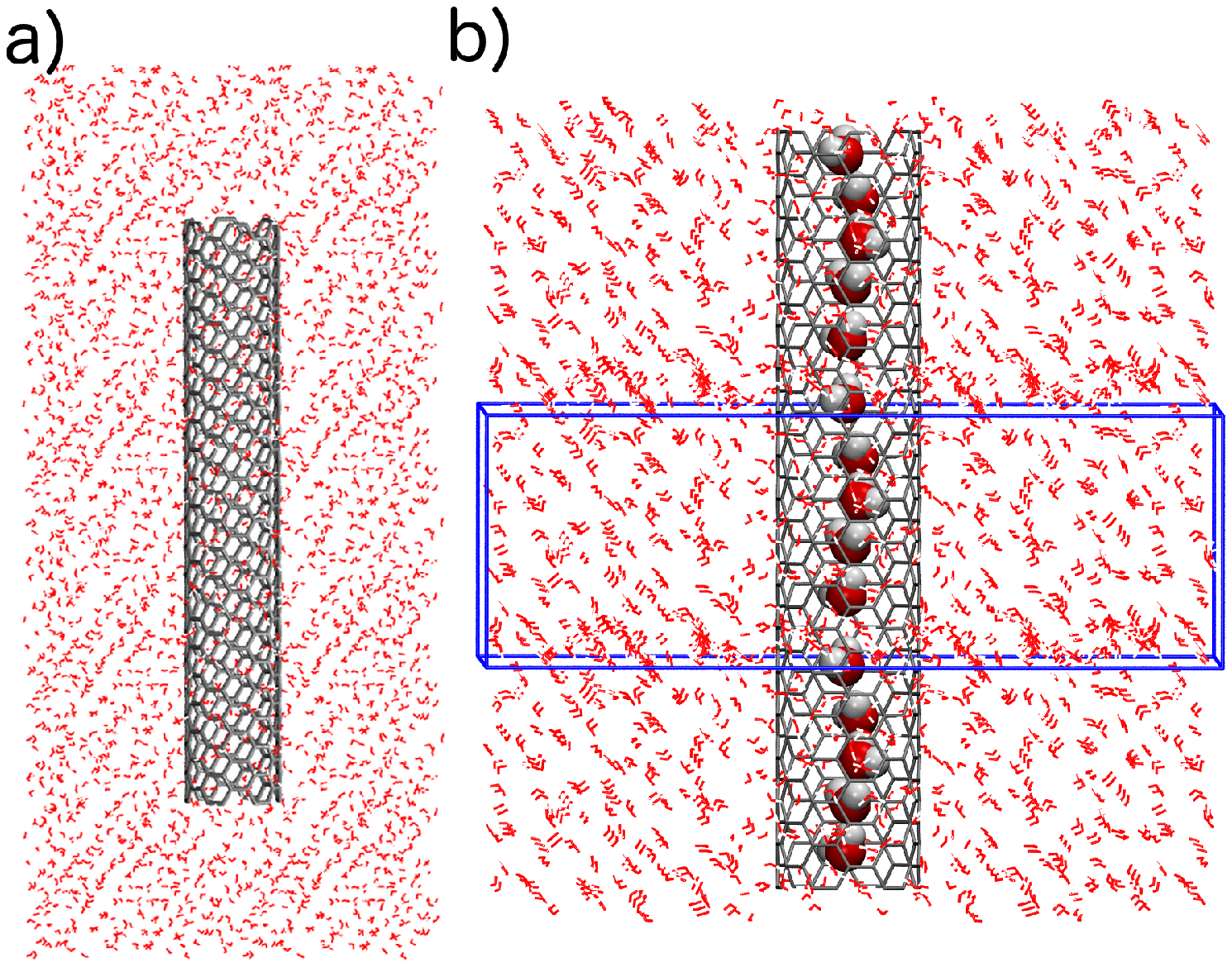}
\caption{Simulated system. Left: $54$~\AA~ long open ended single-wall carbon nanotube. Right: $13.4$~\AA~ long periodically replicated nanotube. 
Confined water molecules are shown in vdW representation.}
\label{simu}
\end{figure}
To make direct comparison with recent studies~\cite{waghe2012}, we performed energy calculations for a $13.4\,$~\AA~ long SWCNT solvated in bulk water ($\sim$ 3000 water molecules). Due to unfavorable interaction with the SWCNT for $\lambda<1$ values, water molecules do not stay long enough inside the nanotube to produce sufficiently long trajectories to give converging velocity autocorrelation function data, making 2PT computations inaccurate. To overcome this problem, all the calculation for different $\lambda$ values were done for water molecules confined inside a periodically replicated nanotube solvated in bulk water, which does not allow the water molecules to escape due to the presence of periodic images of the nanotube at both ends. In all cases, the nanotube was held fixed for the entire duration of simulation.

Entropy and energy calculations were performed for occupancy $N_w=5$ which has been shown to be the minimum free-energy state for a $13.4$~\AA~long nanotube~\cite{waghe2012}. 
To obtain the energy for different components of the system, we partition the potential energy into a sum over atoms~\cite{decomp}. This is done by assigning equal portion of the energy to all participating atoms for vdW energy, bond energy, angle energy and dihedral energy. The per atom energy break up for the long-range Coulomb interaction part is accomplished by calculating the electrostatic potential at each charge site due to all other charges
and multiplying the potential by the charge at that site.

All the simulations were performed in the NPT ensemble at a constant pressure of 1 atmosphere and temperatures ranging from 273 K to 353 K. Constant temperature and pressure were maintained using the Berendsen weak-coupling method as implemented in the PMEMD module of AMBER12~\cite{amber}, with coupling constants of 2.0 ps and 2.0 ps, respectively, for temperature and pressure bath coupling. Long-range electrostatic interactions were computed using the particle-mesh Ewald summation scheme  with a real-space cut-off of $10.5\,$~\AA. LJ interactions were computed using the same cut-off distance. The SWCNT was solvated with at least $15$~\AA~ thick water shell in all 3 directions for all simulations except for the periodically replicated case in which there was no solvation shell in the $z$ direction and both ends of the nanotube were connected across the boundary (see Figure~\ref{simu}).\\

After equilibration for 5 ns, coordinates and velocities of all water molecules were saved for 40 ps long trajectories with 4 fs saving frequency with integration time step of 1 fs. The density of states for entropy computation was calculated from  the Fourier transform of the velocity autocorrelation function of water molecules. Translational and rotational components of the entropy were obtained by decomposing the total velocity into the center of mass velocity and the angular velocity. To get better statistical data, all calculations were performed for four independent trajectories and averaged values were obtained for different physical quantities of interest.\\

\section{Results}
Filling is governed by the difference between the chemical potentials of bulk and confined water. The equilibrium between bulk water outside the nanotube and confined water inside can be altered  by changing either of these chemical potentials. The chemical potential of bulk water can be changed by changing the temperature and the pressure or by adding salts. Changing the interaction strength between water and the carbon atoms that form the surface of the nanotube can also change the equilibrium between bulk and confined water. We have studied the effects of temperature and carbon-water interaction on the filling behavior by computing the free-energy difference between confined and bulk water for different values of these quantities.

\subsection{Effects of Changing the Temperature}
\label{temp}
\begin{figure}
\centering
\includegraphics[scale=0.5]{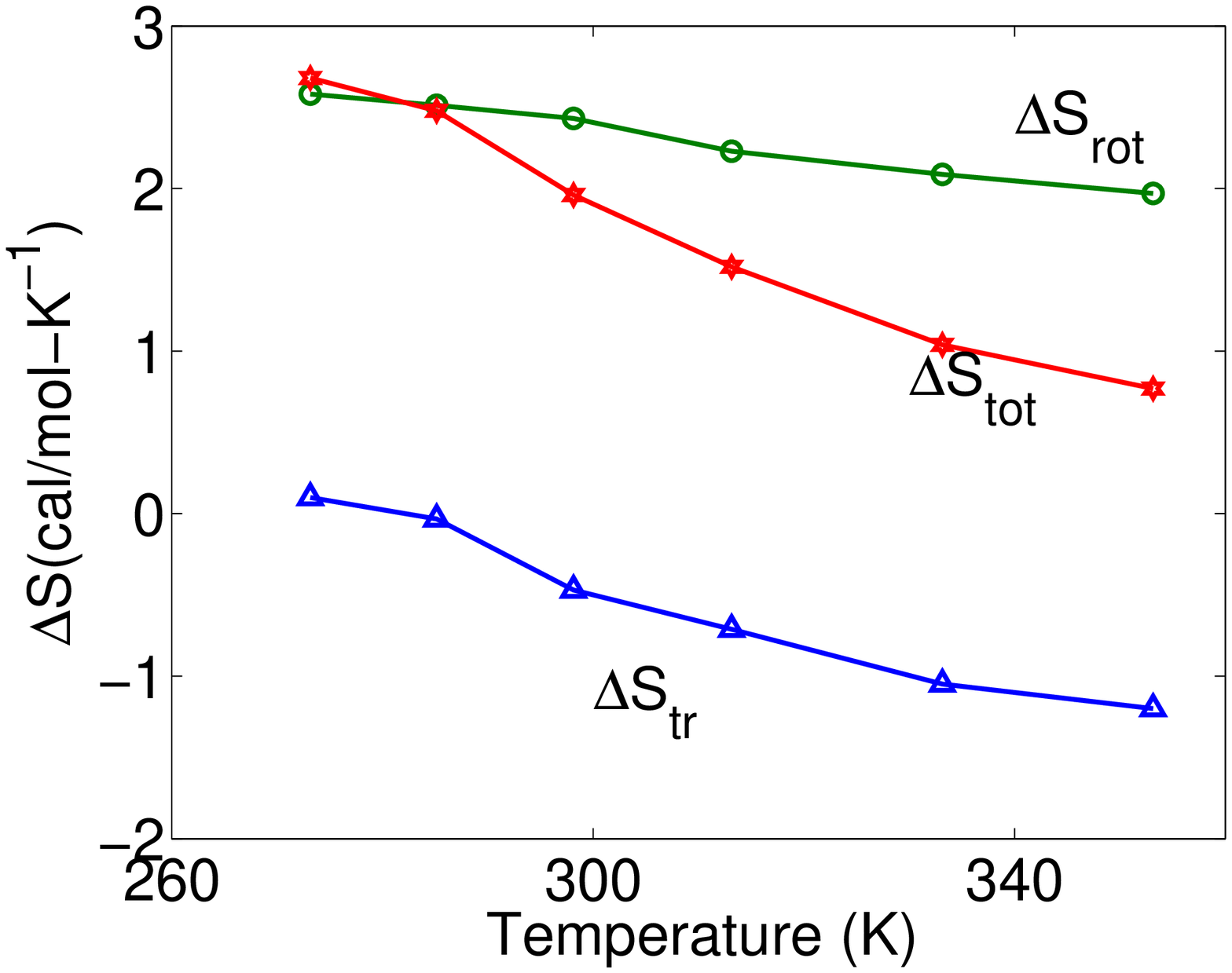}
\caption{Variation of the translational, rotational and total entropies of  transfer per water molecule ($\Delta S =S_{conf}-S_{bulk}$) with the temperature for a $54$~\AA~long open ended nanotube. Both translational and rotational entropies of transfer decrease with increasing temperature.}
\label{ent_temp1}
\end{figure}
\begin{figure}
\centering
\includegraphics[scale=0.5]{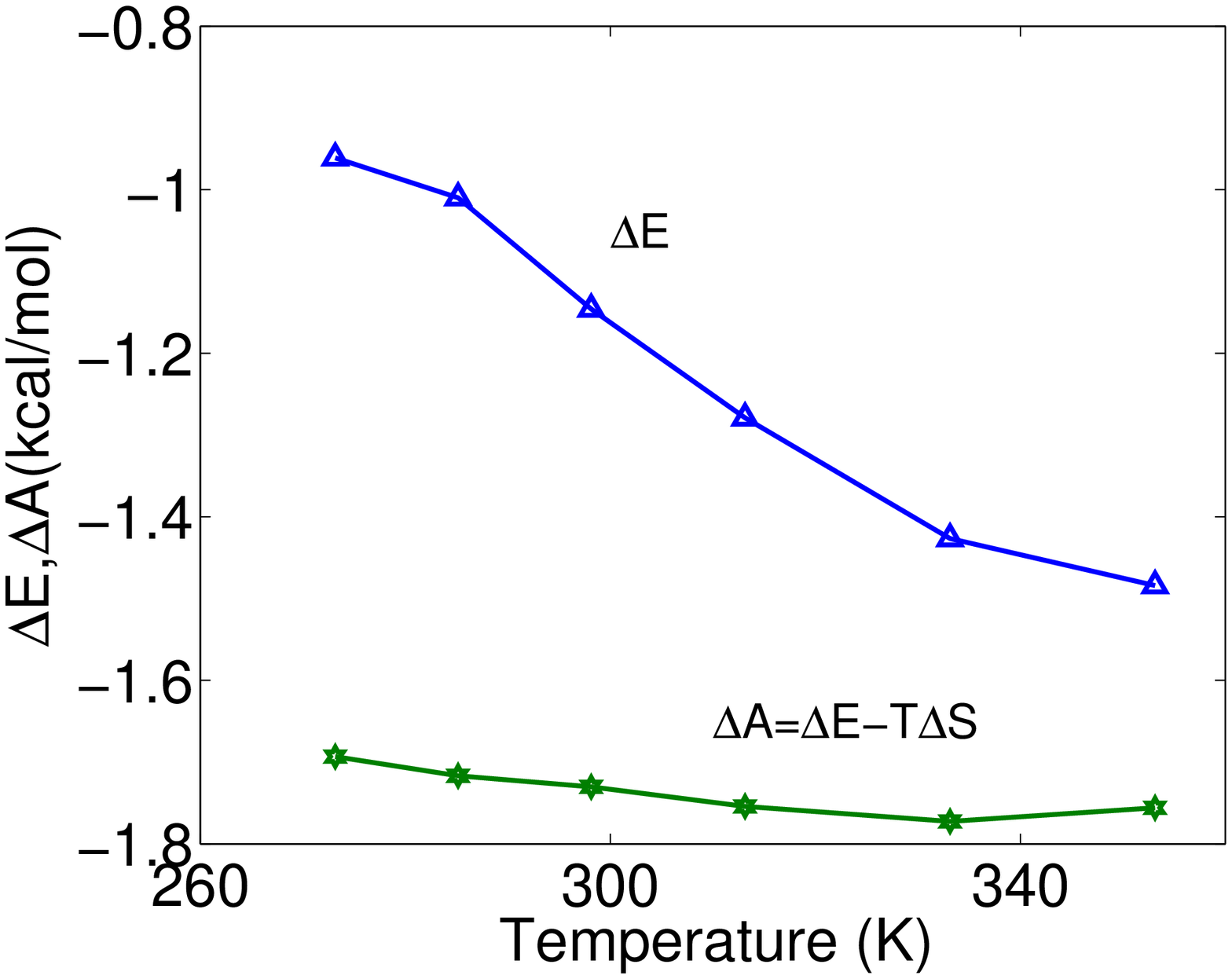}
\caption{Variation of the energy of transfer per water molecule ($\Delta E =E_{conf}-E_{bulk}$) and the free energy of transfer per water molecule,  $\Delta A$, with the temperature for a $54$~\AA~long open ended nanotube.} 

\label{ene_temp1}
\end{figure}
We have computed the entropy and energy of confined and bulk water molecules at different temperatures ranging from 273 K to 353 K. Figure~\ref{ent_temp1} shows the variation of the translational and the rotational entropy of transfer ($\Delta S=S_{conf}-S_{bulk}$) at different temperatures. The translational entropy of confined water is lower than that of bulk water molecules for all the temperatures considered here. In contrast, the rotational entropy of confined water is always higher than that of bulk water over the temperature range considered here. However $\Delta S_{rot}$ decreases with increasing temperature. This can be understood by considering hydrogen bond breaking in bulk water at elevated temperatures. As the temperature increases, more hydrogen bonds break and bulk water molecules acquire relatively more rotational freedom and hence gain rotational entropy. On the other hand confined water remains in the structurally same state with one hydrogen free and the other hydrogen bonded to one of the nearest water molecules in the chain. Hence, the difference between the rotational entropies of confined and bulk water molecules decreases with increasing temperature.

The energy of transfer also exhibits a systematic variation with increasing temperature. Water molecules confined inside the nanotube have lower energy and the energy of transfer defined as $\Delta E = E_{conf}-E_{bulk}$ is a negative quantity. As the temperature is increased, the energy of transfer decreases (increases in magnitude), as shown in Figure~\ref{ene_temp1}. This behavior can again be understood in terms of hydrogen bond breaking at elevated temperatures. In bulk water, hydrogen bonds can be broken more easily at higher temperatures due to increased kinetic energy. However water molecules inside the nanotube are less affected by thermal fluctuations due to strong confinement and a smaller  number of hydrogen bonds are broken as the temperature is increased. Upon increasing  the temperature from 298 K to 333 K, the average number of hydrogen bonds per water molecule in the bulk changes from 3.55 to 2.68 while for the water molecules in confinement, it changes from 1.80 to 1.56. Hence the change in energy of the confined water molecules with increasing temperature is less as compared to that of the bulk water molecules. This causes the energy of transfer to decrease with increasing temperature.\\

Combining the entropy of transfer with the energy of transfer, the free energy of transfer $\Delta A = \Delta E-T\Delta S$ is obtained. The temperature dependence of $\Delta A$ has been shown in Figure~\ref{ene_temp1}. This quantity is negative, consistent with the observation that the nanotube remains filled during the simulation (see Table~\ref{occ}). 
It should be noted that these calculations were done for an open-ended nanotube immersed in a water bath which allows free exchange of confined water molecules with those in the bulk. Hence the average occupancy of water molecules inside the nanotube changes slightly as the temperature is changed (see Table~\ref{occ}).
\begin{table}[htbp]
\centering
\caption{ Energy, entropy and free energy of transfer per water molecule and average occupancy of the nanotube for different temperatures for a $54\,$~\AA~long open-ended nanotube.}
\begin{tabular}{ccccc}\\ \hline \hline
$Temperature$ (K)& $\Delta E$~(kcal/mol)&$T\Delta S$~(kcal/mol)&$\Delta A$~(kcal/mol.)&  Avg. Occupancy\\ \hline
$273.16$&$-0.96\pm0.15$&$0.73\pm0.11$&$ -1.69\pm0.26$ & $19.7$\\ 
$285.15$&$-1.01\pm0.15$&$0.71\pm0.11$&$ -1.72\pm0.26$ & $19.4$\\
$298.15$&$-1.15\pm0.15$&$0.58\pm0.14$&$ -1.73\pm0.29$ & $19.1$\\ 
$313.15$&$-1.28\pm0.15$&$0.47\pm0.14$&$ -1.75\pm0.29$ & $18.9$\\
$333.15$&$-1.43\pm0.15$&$0.35\pm0.14$&$ -1.77\pm0.29$ & $18.7$\\ 
$353.15$&$-1.48\pm0.16$&$0.27\pm0.14$&$ -1.76\pm0.30$ & $18.3$\\ \hline\hline

\end{tabular}
\label{occ}
\end{table}
\subsection{Effects of Changing the Carbon-oxygen Interaction}
\label{vdw}
\begin{figure}
\centering
\includegraphics[scale=0.5]{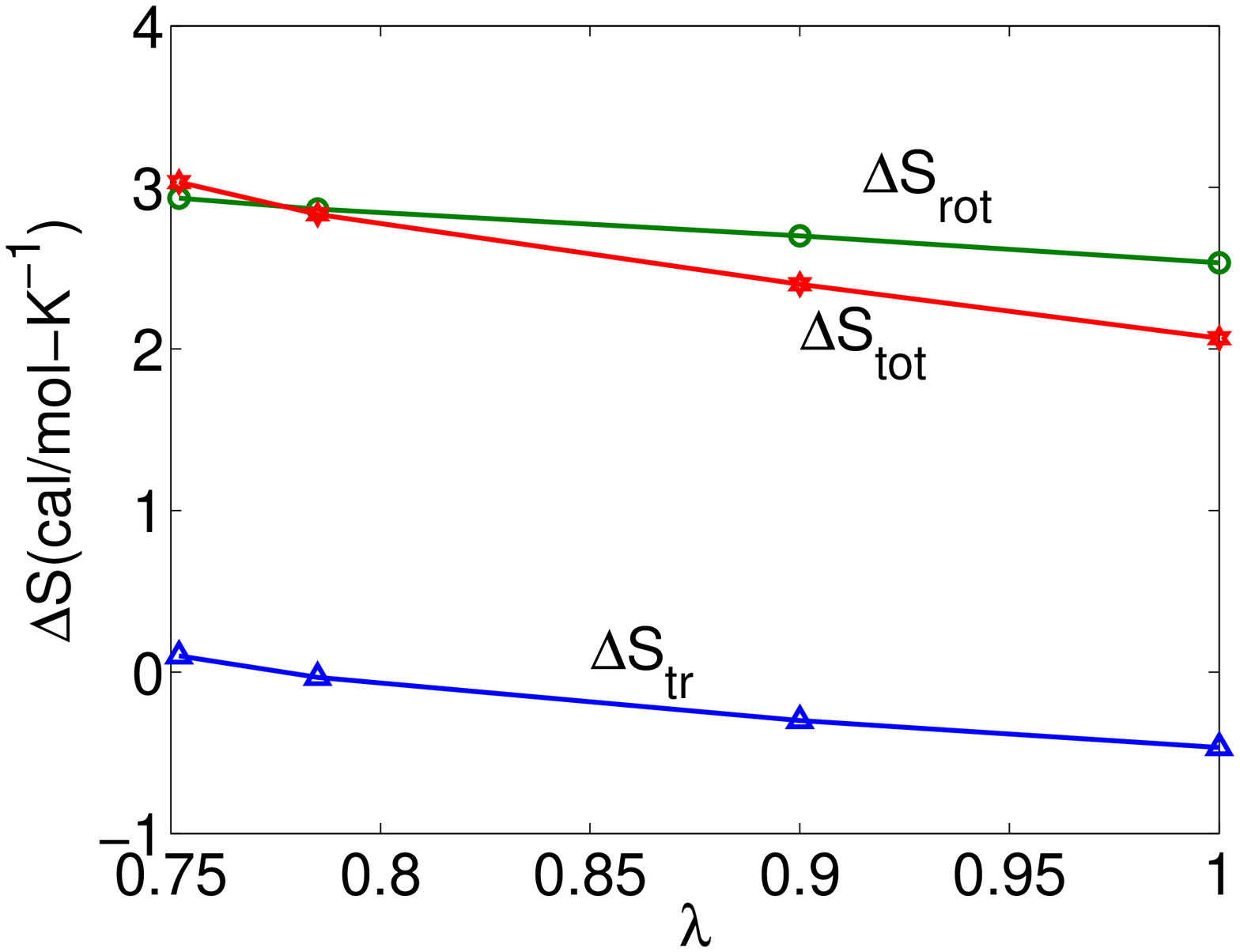}
\caption{Variation of the translational, rotational and total entropy of transfer per water molecule $\Delta S$ as a function of $\lambda$ at T = 300 K for a $13.4$~\AA~ long periodic nanotube with occupancy of 5 water molecules.}
\label{ent_lambda}
\end{figure}

Filling of the carbon nanotube cavity by water molecules is extremely sensitive to the wall-water (carbon-oxygen) interaction strength. Hummer {\it et al.}~\cite{hummer_nature} showed that reducing the depth of the carbon-oxygen LJ potential well by 0.05 kcal/mol results in the drying of the nanotube cavity for significant periods of time. We have studied the thermodynamics of filling systematically as a function of the parameter $\lambda$ in Eq.(\ref{ljmod}) which modifies the oxygen-carbon interaction strength. 
As shown in Figure~\ref{ent_lambda}, both translational and rotational entropies of confined water increase with decreasing $\lambda$, making the nanotube cavity entropically more favorable. An increase in the entropy (both translational and rotational) with decreasing the strength of the attractive part of the interaction, as shown in Figure~\ref{ent_lambda} and Table~\ref{fene_transfer}, is physically reasonable. This observation is contrary to the results of Waghe {\it et al.}~\cite{waghe2012}, who found the entropy of transfer to be independent of $\lambda$.

\begin{figure}
\centering
\includegraphics[scale=0.5]{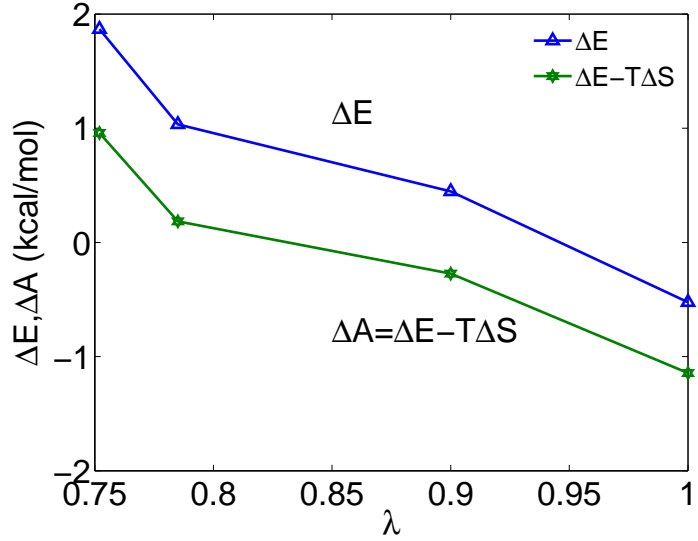}
\caption{Variation of the energy and the free energy of transfer per water molecule, $\Delta E$~ and $\Delta A=\Delta E- T\Delta S$~ respectively, as a function of $\lambda$ at T=300 K for a $13.4$~\AA~ long periodic nanotube with occupancy of 5 water molecules.}
\label{ene_lambda}
\end{figure}

The interaction energy of confined water molecules for different $\lambda$ values has been shown in Table~\ref{fene_transfer}. As expected, scaling down the attractive part of the Lennard-Jones potential makes the nanotube cavity energetically less favorable and the energy of the water molecules inside the nanotube increases (become less negative) with decreasing $\lambda$. The free energy of transfer shows the same trend, as shown in Figure~\ref{ene_lambda}. While the energy of transfer makes the cavity unfavorable for filling, the entropy of transfer makes it favorable (see Figure~\ref{ent_lambda}). These two effects compete with each other to determine the free energy of transfer. The change in energy dominates over the change in entropy, and as shown in Table~\ref{fene_transfer}, the free energy of transfer becomes more unfavorable with decreasing $\lambda$.\\
\begin{table}[t]
\centering
\caption{Energy, entropy and free energy per confined water molecule inside a $13.4\,$~\AA~ long periodic nanotube with occupancy of 5 water molecules for different values of the scaling factor $\lambda$ at T = 300 K.  The values for bulk water are also shown for comparison.}
\begin{tabular}{ccccc}\\ \hline \hline
$\lambda$&$E$~(kcal/mol)&$ TS_{trans}$~(kcal/mol)&$TS_{rot}$~(kcal/mol)&$E-TS$~(kcal/mol)\\ \hline
$1.000$&$ -9.98 \pm (0.16)$ & $3.88\pm(.20)$ & $1.69 \pm (0.15)$ & $-15.55 \pm (0.51)$\\ 
$0.900$&$ -9.00 \pm (0.16)$ & $3.93\pm(.20)$ & $1.74 \pm (0.15)$ & $-14.67 \pm (0.51)$\\
$0.785$&$ -8.42 \pm (0.16)$ & $4.01\pm(.20)$ & $1.79 \pm (0.15)$ & $-14.22 \pm (0.51)$\\ 
$0.752$&$ -7.58 \pm (0.16)$ & $4.05\pm(.20)$ & $1.83 \pm (0.15)$ & $-13.46 \pm (0.51)$\\
$BULK$&$  -9.45 \pm (0.02)$ & $4.02\pm(.05)$ & $0.93 \pm (0.01)$ & $-14.40 \pm (0.08)$\\
 \hline\hline

\end{tabular}
\label{fene_transfer}
\end{table}

\section{Discussion and Conclusion}
\label{discuss}
When water molecules enter the nanotube, they lose hydrogen bonds, which increases their energy, whereas the vdW interaction with the carbon atoms in the nanotube decreases their energy. Reduced carbon-oxygen interaction makes the nanotube cavity unfavorable for water molecules and the nanotube remains empty for a considerable fraction of simulation time. In this case, a good estimate of the occupation probabilities, including the probability $p(0)$~of having no water molecule inside the nanotube, can be made, which is not possible for the actual carbon-oxygen interaction. This allows the use of another method to compute the free energy of transfer for filling. Logarithms of the ratio of the probabilities, $p(N)/p(0)$, give estimates of the free energy of transfer $\Delta A_N$ for various occupancies. Waghe {\it et al.}~\cite{waghe2012} computed the occupation probabilities $p(N)$ ~from simulations of water inside a $13.4$~\AA~ long nanotube of $8.1$~\AA~ diameter with reduced carbon-oxygen interaction and extrapolated the results to the actual carbon-oxygen interaction. From these results, they concluded that it is energetically favorable, but entropically unfavorable for water molecules to be inside the
nanotube, and hence the entry of water molecules inside a (6,6) SWCNT is energy driven. 
Recently Garate {\it et al.}~\cite{garate}, using the thermodynamic integration method for a $31.9\,$~\AA~long (6,6) SWCNT, have shown that the entry of water is entropy driven for lower occupancies, but for the fully occupied state both energy and entropy favor confinement. Favorable energy and entropy 
the fully occupied state of the nanotube is consistent with our result presented here, for both open ended (see Figures 2,3) and periodic nanotubes (see Figures 4,5). However, these results are contrary to those of our previous study~\cite{hemant_jcp} where we found that the confined water molecules have favorable entropy but unfavorable energy inside the nanotube. 
\begin{table}[t]
\centering
\caption{Comparison of the energy, entropy and free energy of transfer per water molecule for a (6,6) nanotube from various studies. The values are given for open ended nanotube only.}
\begin{tabular}{cccc}\\ \hline \hline
Source&$\Delta E$~(kcal/mol)&$T\Delta S$~(kcal/mol)&$\Delta A$~(kcal/mol)\\ \hline
Pascal {\it et al.}\cite{pascal_pnas}&$2.13$&$3.06$&$-0.93$\\
Garate {\it et al.}\cite{garate}&$-1.21$&$0.43$&$-1.63$\\
Waghe  {\it et al.}\cite{waghe2012}&$-2.29$&$-0.76$&$-1.53$\\
This work                          &$-1.15$&$0.58$&$-1.73$\\
 \hline\hline
\end{tabular}
\label{fene_comp}
\end{table}
{\it This difference lies in the method employed in computing the free energy of transfer.   The method used by Waghe {\it et al.}~\cite{waghe2012} and
Garate {\it et al.}~\cite{garate} gives the total free energy of transfer of the whole system in going from zero occupancy to an occupancy state with N water molecules. In contrast the  the free energy of transfer  in our earlier work was computed as  the difference in the energy and entropy of bulk and confined water molecules only and neglects the change of nanotube energy due to filling.}

 For a $54$~\AA~ long SWCNT at 300 K, the change in the nanotube energy due to filling in the fully occupied  state is $\sim -2.04$ kcal/mol per water molecule. The interaction energy  per confined water molecule inside a $54$~\AA~ long nanotube at $298$~K is  $-8.68\,$~kcal/mol without including the change in the nanotube energy due to filling. The interaction energy of water molecules ``only" is higher than the bulk water molecules energy, $-9.57\,$~kcal/mol, at the same temperature. However, upon including the nanotube energy change due to filling, the energy of a confined water molecule becomes $-10.72\,$~kcal/mol indicating a favorable energy of transfer of $-1.15\,$~kcal/mol.\\
  This value qualitatively agrees with the values reported by Waghe {\it et al.}~\cite{waghe2012}($\sim$-2.29 kcal/mol) and Garate {\it et al.}~\cite{garate}($\sim$-1.21 kcal/mol) (Table~\ref{fene_comp}) for open ended nanotube.  
The quantitative differences might be due to different filling states, different nanotube lengths or different water models used. The energy of confined water molecules reported in this study includes the nanotube energy change per water molecule in all cases. The decomposition of the water molecules' energy and nanotube energy due to filling for all cases is given in the supporting information. We also observe that for partially occupied nanotubes, the energy of the confined water molecules is higher than the energy in the bulk even after including the nanotube energy change. As the occupancy increases, the energy of the confined water molecules becomes more favorable, but the entropy of transfer becomes less favorable. This behavior is consistent with that observed by Garate {\it et al.} ~\cite{garate} who used the
thermodynamics integration method to compute the transfer free energy, energy and entropy. 
Results for different occupancies for a $13.4$~\AA~ long periodic SWCNT are presented in the 
supporting information.\\

A comparison of the results for the energy, entropy and free energy of transfer obtained in different studies is shown in Table~\ref{fene_comp}. The
results for $\Delta A$ obtained in different studies are similar, all indicating a favorable free energy of transfer, consistent with the observation
that a nanotube immersed in a bath of water gets spontaneously filled. The values of $\Delta E$ and $\Delta S$ obtained in this work are close to
those reported by Garate {\it et al.} ~\cite{garate}.
Waghe {\it et al.}~\cite{waghe2012} also decomposed the free energy of transfer into entropic and enthalpic components. Assuming that the energy of transfer $\Delta E_N$ and the entropy of transfer $\Delta S_N$ in the expression  $\Delta A_N = \Delta E_N -T \Delta S_N$ are constant over the temperature range studied, a linear fit to $\Delta A_N$ vs. $T$ data was used to extract the values of $\Delta E_N $ and  $\Delta S_N$ from the intercept and slope, respectively. The resulting values of $\Delta E_N $ and  $\Delta S_N$ for reduced carbon-water interaction were extrapolated to $\lambda=1$ to obtain the values of these quantities for real carbon-water interaction. It was further assumed that the entropy of transfer is independent of carbon-water interaction and the energy of transfer was found to exhibit a linear dependence on the scaling factor $\lambda$ over the studied range. Based on these assumptions, Waghe {\it et al.}~\cite{waghe2012} argued that the confined water molecules have lower entropy as compared to bulk water molecules, in disagreement with the findings of this study and previous studies  ~\cite{hemant_jcp,pascal_pnas}. Results of our explicit calculations, presented in previous sections, show that the assumption of Waghe {\it et al.} about temperature independence of $\Delta E$ and $\Delta S$ in the temperature range studied  does not hold true. This may be the reason for the discrepancy between the results and conclusions of Waghe ~\textit{et al.} and those obtained in other studies~\cite{hemant_jcp,pascal_pnas, garate}. The reason for the differences between the results of Pascal {\it et al.}~\cite{pascal_pnas} and those obtained in other studies remains to be explained.{\it It should be noted, that the values of the energy and entropy of transfer are
sensitive to the nanotube length and care should be taken to use same nanotube length while comparing the values presented here.  For  nanotubes of other diameters, both the energy and the entropy of transfer show systematic change as shown in the previous studies ~\cite{pascal_pnas,hemant_jcp,garate}.}\\

In summary, we have used fully atomistic MD simulations to study the effects of changing the temperature and the water-carbon interaction on the thermodynamics of the entry of water molecules in narrow carbon nanotubes. Both entropy and energy of transfer decrease with increasing temperature, keeping the free energy of transfer nearly constant. The entropy and the energy of water molecules inside the nanotube increase as we scale down the attractive part of the LJ interaction between the carbon atoms of the nanotube and the oxygen atoms of water, with the energy increasing faster than the entropy. The change in the nanotube energy due to filling is found to be important for understanding the differences between the results of different studies of the driving forces for water entry into the nanotube. The results presented here will help in reconciling different views on the thermodynamics of water entry inside the hydrophobic channel of narrow carbon nanotubes.
\bibliography{manuscript}
\bibliographystyle{ieeetr}
\begin{acknowledgments}
We thank Shiang-Tai Lin and Tod Pascal for valuable suggestions and comments. We acknowledge DST, India for financial support. HK would like to thank University Grants commission, India for Senior Research Fellowship.
\end{acknowledgments}
\end{document}